\documentclass[english,twocolumn]{article}

\ifx\directlua\undefined\ifx\XeTeXcharclass\undefined
  \usepackage[utf8]{inputenc}                           
  \else\RequirePackage[no-math]{fontspec}[2017/03/31]\fi 
  \else\RequirePackage[no-math]{fontspec}[2017/03/31]\fi 
\usepackage[sort&compress,square,numbers]{natbib}
\usepackage[big,online]{dgruyter}
\usepackage{graphicx}
\usepackage{bm}

\usepackage[utf8]{inputenc}
\usepackage[T1]{fontenc}
\usepackage{mathptmx}
\usepackage{etoolbox}
\usepackage{siunitx}
\usepackage{amsmath,amssymb,amsfonts}%
\usepackage{amsthm}%
\usepackage{mathrsfs}%
\usepackage{hyperref}
\usepackage{xcolor}
\usepackage{nameref}

\theoremstyle{dgthm}

\theoremstyle{dgdef}

\begin{document}

\articletype{Research Article}

\author*[1]{Oliver Kuster}
\author[2]{Yannick Augenstein}
\author[3]{Roberto Narváez Hernández}
\author[4]{Carsten Rockstuhl}
\author[5]{Thomas Jebb Sturges}
\affil[1]{Institute of Theoretical Solid State Physics, Karlsruhe Institute of Technology (KIT), Kaiserstrasse 12, 76131 Karlsruhe, Germany, oliver.kuster@kit.edu; https://orcid.org/0009-0003-2717-7690}
\affil[2]{Flexcompute Inc, Belmont, MA, USA}
\affil[3]{Institute of Theoretical Solid State Physics, Karlsruhe Institute of Technology (KIT), Kaiserstrasse 12, 76131 Karlsruhe, Germany}
\affil[4]{Institute of Theoretical Solid State Physics, Karlsruhe Institute of Technology (KIT), Kaiserstrasse 12, 76131 Karlsruhe, Germany and Institute of Nanotechnology, Karlsruhe Institute of Technology (KIT), Kaiserstrasse 12, 76131 Karlsruhe, Germany; https://orcid.org/0000-0002-5868-0526}
\affil[5]{Institute of Theoretical Solid State Physics, Karlsruhe Institute of Technology (KIT), Kaiserstrasse 12, 76131 Karlsruhe, Germany; }
\title{Inverse Design of 3D Nanophotonic Devices with Structural Integrity using Auxiliary Thermal Solvers }
\runningtitle{Inverse Design of 3D Nanophotonic Devices with Structural Integrity using Auxiliary Thermal Solvers }
\abstract{3D additive manufacturing enables the fabrication of nanophotonic structures with subwavelength features that control light across macroscopic scales. Gradient-based optimization offers an efficient approach to design these complex and non-intuitive structures. However, expanding this methodology from 2D to 3D introduces complexities, such as the need for structural integrity and connectivity. This work introduces a multi-objective optimization method to address these challenges in 3D nanophotonic designs.
Our method combines electromagnetic simulations with an auxiliary heat-diffusion solver to ensure continuous material and void connectivity. By modeling material regions as heat sources and boundaries as heat sinks, we optimize the structure to minimize the total temperature, thereby penalizing disconnected regions that cannot dissipate thermal loads. Alongside the optical response, this heat metric becomes part of our objective function. We demonstrate the utility of our algorithm by designing two 3D nanophotonic devices. The first is a focusing element. The second is a waveguide junction, which connects two incoming waveguides for two different wavelengths into two outgoing waveguides, which are rotated by 90° to the incoming waveguides. Our approach offers a design pipeline that generates digital blueprints for fabricable nanophotonic materials, paving the way for practical 3D nanoprinting applications.}
\journalname{Nanophotonics}
\journalyear{2025}
\journalvolume{aop}

\maketitle

\vspace*{-6pt}
\section{Introduction}
3D nanoprinting enables the fabrication of photonic devices with voxel sizes below the wavelength of light, achieving intricate structures that seamlessly integrate features from the nanoscale to millimeter or even centimeter scale \cite{fischerThreedimensionalOpticalLaser2013, barner-kowollik3DLaserMicro2017, hahn2020rapid, kiefer2024multi, somers2024physics}. Fabrication happens upon two-photon absorption in a photoresist within a tiny volume into which a writing laser is focused. Two-photon absorption triggers the polymerization of the photoresist, and a structure is formed. Micro-optical components are widely explored with such fabrication techniques \cite{10.1063/1.5064401, Siegle:24, ruchka20243dprintedaxiconenablesextended, schmid20233d, papamakarios2024cross, tsilipakos2024polymeric, doi:10.1021/acsphotonics.4c01179, doi:10.1021/acsphotonics.4c00524, doi:10.1126/science.abq3037, dana2024free, https://doi.org/10.1002/adfm.202211280, doi:10.1021/acsphotonics.4c00953, 10636389}, and complex photonic systems with a non-intuitive design have also been realized \cite{https://doi.org/10.1002/adma.202310100, bürger2024onchiptwistedhollowcorelight, wang2024two, https://doi.org/10.1002/admt.202300893, weinacker2024multi}. By raster scanning the laser across a volume, free-form designs in three dimensions can be realized, where every voxel constitutes a design parameter, {\it i.e.}, it may or may not be polymerized. In principle, this high dimensionality promises structures with almost arbitrary functionalities due to the many degrees of freedom available.

However, the high dimensionality of the parameter space is simultaneously a blessing and a curse. All degrees of freedom must be efficiently adjusted while accommodating fabrication and application constraints in the design. Due to the complexity of such devices, efficient numerical methods are indispensable to solve the inverse problem \cite{christiansenInverseDesignPhotonics2021, moleskyInverseDesignNanophotonics2018, noh2023inverse, so2020deep, wiecha2021deep, radford2024inversedesignunitarytransmission, Dai:22, https://doi.org/10.1002/adma.202305254, doi:10.1021/acsphotonics.2c01006}. For free-form 3D nanoprinting, global optimization techniques are no longer applicable, and only gradient-based inverse design methods can be used. Here, the adjoint formalism is particularly beneficial as it allows us to iteratively update a design with only two simulations of our system to show improved performance in relation to a chosen objective function. The adjoint formalism is the core of topology optimization, which is a highly efficient method in designing nanophotonic devices \cite{jensenTopologyOptimizationNanophotonics2011, bendsoeBibliographicalNotes2004, sapra2019inverse, muraiMultiscaleTopologyOptimization2023, piggottInverseDesignDemonstration2015, Nanda:24, Hassan:22, gedeon2025time, igoshin2024inversedesignmieresonators, 10618079, jin2018inverse}.

\begin{figure*}
    \centering
    \includegraphics[width=\linewidth]{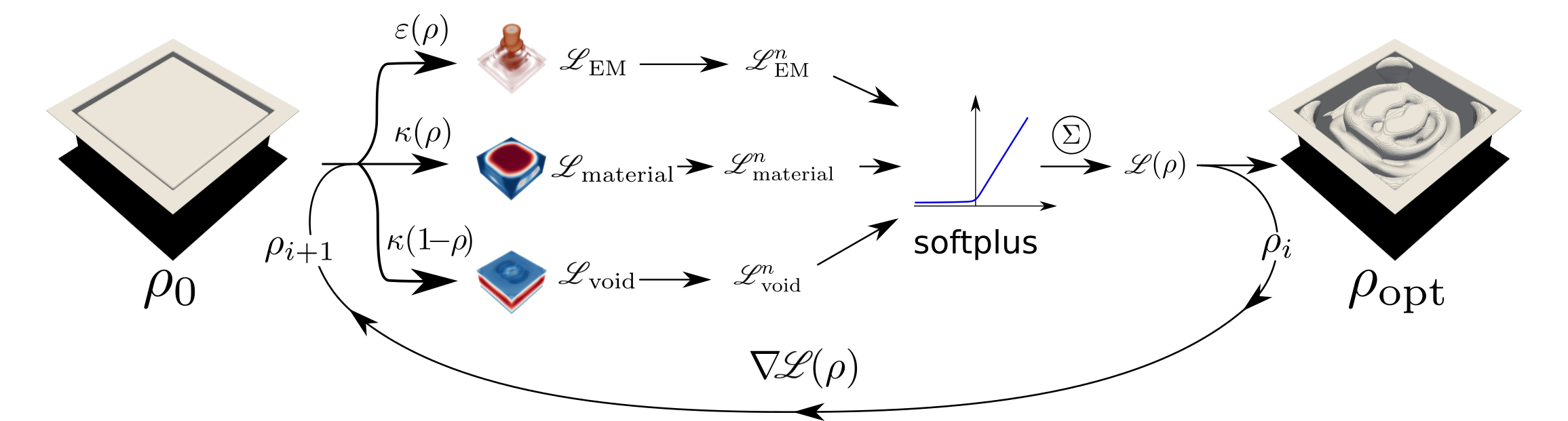}
    \caption{Sketch of our design procedure. The initial design, expressed in terms of a material density $\rho \in [0,1]$, is mapped to three different material distributions. Firstly, it is mapped to the permittivity $\epsilon(\rho)$ of the dielectric material to be printed, from which we calculate the electromagnetic response. Additionally, a fictitious heat-diffusion through the system is studied in two dedicated auxiliary simulations by assuming the written material and the void as heat sources, respectively. The material density is mapped to a thermal conductivity $\kappa(\rho)$ for the material, and $\kappa(1-\rho)$ for the void. From all three simulations, an objective function is evaluated. Each sub-objective is then renormalized, passed through a \textit{softplus}-function, and summed to the final objective function. In combination, these sub-objectives aim to balance the goal of maximizing a given optical figure of merit while ensuring structural connectivity of the devices with no cavities. The \textit{softplus}-function ensures that the fictitious thermal performance of the design stops being optimized once structural connectivity is achieved. The gradients of the objective function relative to the material density at every point in the design space are then used to update the design iteratively and to optimize the digital blueprint with structural integrity.}
    \label{fig:enter-label}
\end{figure*}

However, not just the optical functionality matters in the design. Besides the usual fabrication constraints, such as minimum feature size, a prime design challenge is ensuring the structural integrity of the 3D nanoprinted devices. We must ensure the final design is fully connected and does not collapse on itself. In other words, free-floating components are not an option for realistic devices. Simultaneously, we must ensure that the design does not include cavities. Cavities would trap the undeveloped photoresist, which is detrimental to the optical functionality. 

One way to ensure the structural integrity of the material is by using physics-based constraints \cite{augensteinInverseDesignNanophotonic2020, jokisch2024engineeringopticalforcesmaxwell, martinez2024topology}.
In this work, we present a method to enforce the structural integrity of 3D photonic structures by using an auxiliary heat-diffusion solver. Comparable approaches have already been discussed in the past to optimize 2D structures \cite{christiansen2023inverse, liStructuralTopologyOptimization2016}.
While structural solvers exist, we require a framework which is differentiable and can easily be combined with an electromagnetic simulation. The heat solver provides a direct and computationally cheaper approach to enforcing global connectivity, which—at the nanoscale—is typically sufficient to also guarantee mechanical stability.
In addition, the heat based solver allows us to easily enforce connectivity of the void, which is a requirement for the 3D additive manufacturing process of nanophotonic devices. Using structural solvers which rely on vectorial forces \cite{augensteinInverseDesignNanophotonic2020} would become difficult and restrict the design space too much.

To evaluate how well-connected a given structure is, we consider the written structure to be a heat source. By solving a heat-diffusion equation, we then study how well the structure can dissipate its heat to strategically placed heat sinks at the outer rims of the design region. As we permit diffusion only inside the written material, the material must be well-connected to reach a low final temperature. By calculating the total temperature of the device (upon integration of the temperature distribution across the design), we measure how well connected the structure is. This process is repeated for the void, which ensures that no cavities are formed within the design. To be clear, the thermal solver is purely fictitious and is used only within the optimization pipeline to ensure connectivity.

These two separate heat-diffusion problems are solved with a finite-element method. The electromagnetic simulation is solved with a finite-difference frequency-domain (FDFD) method \cite{fischbachJandavidfischbachJaxwell2024}.
For each of these three simulations, an adjoint problem is also solved to enable our multi-objective optimization workflow with gradient-based optimization.
The overall objective function is constructed as the sum of three sub-objectives, each of which is normalized relative to a predefined threshold and processed through a \textit{softplus} function. The normalization ensures balanced contributions from the electromagnetic, material, and void simulations \cite{schubertInverseDesignPhotonic2022} while the \textit{softplus} function deactivates further thermal optimization once structural connectivity is established.
The additional computational cost of the thermal simulations is comparatively low. Still, the outcome guarantees that the resulting devices possess structural integrity. Therefore, these designs provide a digital blueprint for direct fabrication. In addition, we find designs that are almost as optically performant as designs optimized only for their optical response. With this, we develop a critical component of a design pipeline for 3D nanoprinted photonic devices that will find applications in perceiving a future generation of novel structures that can serve societal needs. 

Besides this introduction, the paper is structured into three sections. In the following section, we outline our general design pipeline based on topology optimization. Afterward, in Section \nameref{Results}, we demonstrate the applicability of our approach in two carefully selected devices. Finally, we conclude on our work in Section \nameref{Discussion}.

\section{Topology Optimization} \label{TopOp}
Topology optimization, as used in our design pipeline, is a gradient-based, local optimization method. Given a figure of merit (also called the objective function) $\mathcal{L}({\rho})$, topology optimization can formally be written as
\begin{align}
    \min_{{\rho}}\,\,\,&\mathcal{L}({\rho})\\
    \text{s.t.}\,\,\,&c_i({\rho}) = 0&\nonumber\\
    \text{s.t.}\,\,\,&c_j({\rho}) < 0\,,\nonumber
\end{align}
with equality constraints $c_i({\rho})$ and inequality constraints $c_j({\rho})$, $i, j \in \mathbb{N}$.
By calculating $\nabla_\rho \mathcal{L}(\rho)$ and using gradient-based optimizers, a local minimum for the figure of merit can be found for any initial condition chosen as the starting point for the iterative design process. In our case, $\rho$ represents a material density that can be mapped to either the relative permittivity or the thermal conductivity of our designs. In each case, it represents the material distribution we aim to optimize.

A forward simulation is performed to evaluate the objective function. Then, using the adjoint method, the gradients of the objective function with respect to each degree of freedom can be calculated efficiently at the cost of only one additional simulation, called a backward simulation \cite{jensenTopologyOptimizationNanophotonics2011, lalau-keralyAdjointShapeOptimization2013}. Notably, the computational complexity of calculating the gradients using the adjoint method is independent of the number of degrees of freedom, enabling the free-form design of structures characterized by many millions of parameters.

\subsection{Forward Simulation}
To perform the forward simulation, the material density must first be mapped to the corresponding material distribution. We will optimize the density in the design region $\mathcal{D} \subseteq \mathbb{S} \subseteq \mathbb{R}^3$, which is embedded into the simulation domain $\mathbb{S}$.
The continuous material density is represented as $\rho:=\rho(x, y, z)$, with $\rho(x_0, y_0, z_0) \in [0, 1],\ (x_0, y_0, z_0)\in \mathcal{D}$. Discretizing the spatial distribution of $\rho$ divides the design region into voxels, each of which can then be independently parameterized, enabling free-form structural design \cite{bendsoeOptimalShapeDesign1989, sigmundTopologyOptimizationApproaches2013, zhouMinimumLengthScale2015}.
To map $\rho$ to the physical parameters required for our simulations, we use multiple, continuously differentiable transformations. The transformations map the material density to the electric permittivity or the thermal conductivity, respectively, while also incorporating fabrication limitations.

As such, we first apply a Gaussian filter to $\rho$ to enforce a minimum feature size of $\SI{100}{nm}$ \cite{doi:10.1126/science.abq3037} for our structures.
\begin{equation}
    \Tilde{\rho} = \rho * w(\sigma) \,\,\, \text{with} \,\,\, w(\sigma) = e^{-\frac{x^2 + y^2 + z^2}{2\cdot\sigma^2}}\,.
\end{equation}
The Gaussian filter is given by a discrete convolution of $\rho$ with a Gaussian kernel $w(\sigma)$, with a kernel size of $l=2 \cdot \left \lceil{4 \cdot \sigma}\right \rceil + 1$. The minimum feature size itself can be approximated as $\sqrt{3}\sigma$. Considering our minimum feature size of $\SI{100}{nm}$ and the resolution of our simulations, we use $\sigma\approx \SI{2.3}{px}$.

We are only interested in designs consisting of either ``void'' or ``material'' with no intermediate values. This corresponds to only two electric permittivities or thermal conductivities. The binarization of the material density is imposed by a soft threshold filter \cite{christiansenInverseDesignPhotonics2021, wangProjectionMethodsConvergence2011}
\begin{equation}
    \hat{\Tilde{\rho}} = \frac{\tanh(\beta\alpha) + \tanh(\beta(\Tilde{\rho} - \alpha))}{\tanh(\beta\alpha) + \tanh(\beta(1-\alpha))}\,.
\end{equation}
Here, $\beta$ controls the steepness of the function and quantifies the degree of binarization, while $\alpha$ defines its center. We set $\alpha = 0.5$ in our simulations and gradually increase $\beta$ from $1$ to $30$ throughout the optimization. Initially, $\beta$ is kept low to allow broad exploration of the parameter space. As the optimization progresses, we incrementally increase $\beta$ to enforce binarization, ensuring convergence towards a fabricable device.

The last set of transformations interpolates the binarized density to the physical quantities using a linear mapping.
For the electromagnetic simulation, this is given by
\begin{equation}
    \epsilon_r = \epsilon_\text{min} + \hat{\Tilde{\rho}} (\epsilon_\text{max} - \epsilon_\text{min})\,,
\end{equation}
where $\varepsilon_\text{min}$ and $\varepsilon_\text{max}$ are the electric permittivity of the void and the material, respectively.
For the temperature simulation, we use
\begin{equation}
    \kappa = \kappa_\text{min} + \hat{\Tilde{\rho}} (\kappa_\text{max} - \kappa_\text{min})\,,
\end{equation}
where $\kappa_\text{min}$ and $\kappa_\text{max}$ refer to the thermal conductivity of the void and the material, respectively \cite{augensteinInverseDesignNanophotonic2020}. Please note that the heat conductivities considered are purely fictitious and do not correspond to the actual material that will be printed. We choose to keep the values representing physical properties as constant and use simulation parameters as our hyperparameters for the optimization. The values are considered in the auxiliary heat solver only and are chosen to ensure connectivity among all domains. Once we have the material distributions, we can consider them in the respective simulation. From these simulations, we can judge the suitability of a given device for its anticipated purpose. Two types of simulations are performed: the electromagnetic simulation and the heat-diffusion simulations.   

We simulate the electromagnetic response using an FDFD Maxwell solver \cite{fischbachJandavidfischbachJaxwell2024} and evaluate the total electric field distribution $\vec{E}(x, y, z, \omega)$ at the frequency $\omega$ at every point in space for a given illumination.
We take our design region $\mathcal{D}$ and embed it into the full simulation domain $\mathbb{S}$. The total electric field distribution is then used to calculate the electromagnetic figure of merit $\mathcal{L}_\text{EM}:=\mathcal{L}_\text{EM}(\hat{\Tilde{\rho}})$. For the purpose of readability, throughout this manuscript we omit the explicit dependency of our figures of merit on $\hat{\Tilde{\rho}}$. The electromagnetic figure of merit encodes the desired functionality of the device into a single scalar value that indicates how good the optical performance is. 
Therefore, by finding a local maximum of $\mathcal{L}_\text{EM}$, we get a density distribution of $\hat{\Tilde{\rho}}$ which manipulates the electromagnetic wave in the desired way.

In addition to the electromagnetic simulation, two heat-diffusion simulations are performed, from which we judge the connectivity of the material and the voids.
Specifically, we solve Poisson's equation
\begin{align}\label{eqn:heat}
    -\kappa(\hat{\Tilde{\rho}})\nabla^2 u(x, y, z) &= q(\hat{\Tilde{\rho}})\\
    \text{s.t.}\,\,\, u(x_0, y_0, z_0) &= 0  \,\,\, \text{for } (x_0, y_0, z_0) \in U \subseteq\partial \mathcal{D}\,,\nonumber\\
    \text{s.t.}\,\,\,\nabla u(x_0, y_0, z_0) &= 0 \,\,\, \text{for }(x_0, y_0, z_0) \in \partial \mathcal{D}\backslash U\,,\nonumber
\end{align}
where $u(x, y, z)$ is the scalar temperature distribution of the system, $\kappa(\hat{\Tilde{\rho}})$ the thermal conductivity, and $q(\hat{\Tilde{\rho}})$ the heat generation rate, \textit{i.e.} corresponding to the heat sources. 
A steady-state temperature distribution is obtained by solving equation \ref{eqn:heat} with an in-house finite element solver using an H8 element. For readability purposes, we implicitly assume the boundary conditions to hold true for our optimization problem and do not state them explicitly anymore.
When optimizing for material connectivity, we designate material regions as heat sources, \textit{i.e.} $q(\hat{\Tilde{\rho}}) = \hat{\Tilde{\rho}}$, and place heat sinks at points $U$ along the boundary. We strategically position the heat sinks within regions where material is known to be desirable, like waveguide ports.
We assume Neumann boundary conditions for $\partial \mathcal{D}\backslash U$. Specifying heat sinks will lead to Dirichlet boundary conditions for $U$.
Note that the heat sinks do not need to be on the boundaries but can, in principle, be placed anywhere.
We assume a small background heat conductivity everywhere in space to avoid the temperature diverging for non-connected structures. Such a setting allows us to reach an equilibrium with a high but finite temperature in spatial domains not connected to the sinks. The finite element method requires us to set a non-zero heat conductivity for $\kappa_\text{min}$. Our chosen value range for the thermal conductivity is $\kappa_\text{min}=10^{-5}\,\frac{\text{M}W}{\upmu m\cdot K}$ and $\kappa_\text{max}=1\,\frac{\text{M}W}{\upmu m\cdot K}$.
Structural integrity is achieved by minimizing the total temperature in the system, encouraging the optimizer to connect all material regions to the heat sinks, thereby reducing the temperature within the structure.

The simulation that ensures connectivity of the voids uses $q(\hat{\Tilde{\rho}}) = 1-\hat{\Tilde{\rho}}$ as heat sources, enforcing connectivity of the void and thus avoiding the formation of cavities that trap unwritten photoresist in the final structure. The edges of the design domain that were not considered heat sinks in the material-connectivity simulation will now act as heat sinks in the void-connectivity simulation.

In contrast to the electromagnetic simulation, in the heat-diffusion simulations, we consider only the design region $\mathcal{D}$ and not the full simulation domain. Such a restriction saves computational time and memory, as only $\mathcal{D}$ is relevant for the structural integrity.

Our figure of merit for the thermal simulations is the integrated final temperature distribution within the material (or respectively void). We define these figures of merit as $\mathcal{L}_\text{material} = \mathcal{L}_\text{heat}(q=\hat{\Tilde{\rho}})$ and $\mathcal{L}_\text{void} = \mathcal{L}_\text{heat}(q=1 - \hat{\Tilde{\rho}})$ respectively, where
\begin{equation}
\mathcal{L}_{\text{heat}} = \int_\mathcal{D} u(x, y, z)\,\,\,\text{d}x\text{d}y\text{d}z\,.
\label{eq:heat}
\end{equation}
This implies that we strive to minimize the final temperature within the design region in both simulations.

\subsection{Optimization Problem}
Using the three forward simulations, we get three contributions to our figure of merit, which must be balanced so that the optimizer weights each one appropriately.
Since all these simulations yield vastly different numerical values, we first renormalize each sub-objective to become comparable. 

For $\mathcal{L}_\text{material}$, we define
\begin{equation}
    \mathcal{L}^n_\text{material} = \frac{\mathcal{L}_\text{material} - \mathcal{L}^\text{thresh}_\text{material}}{\mathcal{L}^\text{thresh}_\text{material}}\,,
\end{equation}
to renormalize the contribution from the material heat-diffusion equation.
We introduce a threshold value $\mathcal{L}^\text{thresh}_\text{material}$ where the heat-dissipation is good enough that the material has a high probability of being fully connected. Therefore, the sign of $\mathcal{L}^n_\text{material}$ provides a binary measure of whether the device is connected enough or not. This will allow us to formulate a multi-objective optimization scheme that deprioritizes those sub-objectives that have reached their threshold value. 
In a similar fashion, we renormalize $\mathcal{L}_\text{void}$ to $\mathcal{L}^n_\text{void}$ using a threshold value $\mathcal{L}^\text{thresh}_\text{void}$.

Since we aim to minimize both $\mathcal{L}^n_\text{material}$ and $\mathcal{L}^n_\text{void}$, while the contribution from $\mathcal{L}_\text{EM}$ has to be maximized, $\mathcal{L}_\text{EM}$ is renormalized as
\begin{equation}
    \mathcal{L}^n_\text{EM} =\frac{\mathcal{L}^\text{thresh}_\text{EM} - \mathcal{L}_\text{EM}}{\mathcal{L}^\text{max}_\text{EM} - \mathcal{L}^\text{thresh}_\text{EM}}\,.
\end{equation}
We choose $\mathcal{L}^\text{thresh}_\text{EM}$ to be roughly the same as $\mathcal{L}_\text{EM}$ of a design which was optimized only for its optical performance. Such a choice ensures that we can still find highly efficient designs, even with the more restricted parameter space, since the optimization will mostly focus on $\mathcal{L}_\text{EM}$ once the threshold values for the material and the void have been reached.
We choose $\mathcal{L}^\text{max}_\text{EM}$ to be larger than any $\mathcal{L}_\text{EM}$ during the optimization, so we can ensure that $\mathcal{L}^n_\text{EM}$ stays larger than $-1$ for big values of $\mathcal{L}_\text{EM}$.

As discussed, we chose $\mathcal{L}^n_\text{material}$ and $\mathcal{L}^n_\text{void}$ such that they become negative once the threshold value is reached, which corresponds to fully connected material (or void). As a subsequent step, we apply a \textit{softplus} function to the three sub-objectives that assigns a vanishing weight for negative values,

\begin{equation}
    \text{\textit{softplus}}(\mathcal{L}^n) =
    \begin{cases}
        \ln(1 + e^{\mathcal{L}^n}), & \text{if}\ \mathcal{L}^n < 0.4\\
        \mathcal{L}^n, & \text{otherwise}\, .
    \end{cases}
\end{equation}

We also add a regularization term as a fourth sub-objective, which penalizes non-binarized voxels. This binarization penalty is given as
\begin{align}
    \mathcal{L}_\text{binary} &= \int_\mathcal{D} 4\cdot \hat{\Tilde{\rho}}(x, y, z) \cdot (1 - \hat{\Tilde{\rho}}(x, y, z))\,\,\,\text{d}x\text{d}y\text{d}z\,,
\end{align}
and renormalized to $\mathcal{L}^n_\text{binary} = \frac{\mathcal{L}_\text{binary} - \mathcal{L}^\text{thresh}_\text{binary}}{\mathcal{L}^\text{thresh}_\text{binary}}$ using $\mathcal{L}^\text{thresh}_\text{binary}=10^{-2.3}\,\upmu m^3$. We only turn on the binarization penalty once we are close to convergence. At this point, most of the voxels are binarized anyway. By turning on the binarization penalty, we can binarize the few remaining non-binarized voxels without majorly impacting the properties of the design.

To combine the contributions from each individual figure of merit into a single scalar, we use the $l^2$ norm. Our full optimization problem is then given as
\begin{align}
 &\min_\rho \mathcal{L}(\hat{\Tilde{\rho}}) = \sqrt{\sum_{i} |\text{\textit{softplus}}(\mathcal{L}^n_i(\hat{\Tilde{\rho}}))|^2}\\
  &\text{s.t.}  \,\,\, (\nabla \times \nabla \times - \omega^2\varepsilon(\hat{\Tilde{\rho}}))\vec{E}(\omega) = -i\omega \vec{J}(\omega)\,,\nonumber\\
  &\text{s.t.} \,\,\, -\kappa(\hat{\Tilde{\rho}})\nabla^2 u(x, y, z) = q(\hat{\Tilde{\rho}})\,,\nonumber\\
  &\text{s.t.} \,\,\, -\kappa(1-\hat{\Tilde{\rho}})\nabla^2 u(x, y, z) = q(1-\hat{\Tilde{\rho}})\,,\nonumber\\
 &\text{s.t.} \,\,\, 0 \le\hat{\Tilde{\rho}}(x, y, z) \le 1\,,\nonumber
\end{align}
with $i\in \{\text{EM}, \text{material}, \text{void}, \text{binary}\}$.
We choose to formulate our figure of merit using a \textit{softplus} function instead of simply using prefactors for two reasons.
First, the \textit{softplus} function has a built in cutoff. This means, that once a certain threshold value is crossed, the respective sub-objective will barely be considered in the full figure of merit. Since we are not interested in the thermal performance of our device, the difference between two values of $\mathcal{L}_\text{heat}$ will not make a significant difference in the structural integrity once the connectivity threshold is crossed. This helps us to avoid getting stuck in a local minimum where the thermal performance is optimized but the electromagnetic performance is not.
Second, since we are only interested in crossing certain threshold values for our subobjectives, our figure of merit is consequently less sensitive to the choice of our hyperparameters, avoiding the need for a proper hyperparameter optimization as we discuss later on.

The main disadvantage of this approach is that it requires some prior knowledge about the threshold values. Only with this prior knowledge can a proper renormalization be done. Still, the method is fairly robust. A rough estimate of the threshold values is usually sufficient, saving time compared to a dedicated hyperparameter optimization. The choice of the values for $\mathcal{L}^\text{thresh}_\text{heat}$ mostly depend on the surface the respective heat sinks occupy as well as the overall volume of the structure. If the heat sinks occupy more surface, the system has more efficient means to disperse the heat, making it less prone to the initial value of $\mathcal{L}^n_\text{heat}$. A bigger volume for the structure means more heat sources. Nonetheless, we have found, that choosing $\mathcal{L}^\text{thresh}_\text{heat}$, so that the initial value for $\mathcal{L}^n_\text{heat}$ is in $(0, 1)$ works sufficiently well for most structures. A good initial value tends to be $\mathcal{L}^n_\text{heat}=0.5$.
To get these initial values for $\mathcal{L}^n_\text{heat}$, one heat simulation for the material and one heat simulation for the void on the initial structure are required, which gives us $\mathcal{L}_\text{heat}$ which we can then use to determine $\mathcal{L}^\text{thresh}_\text{heat}$ to give us our desired initial $\mathcal{L}^n_\text{heat}$.

There are various options to choose $\mathcal{L}^\text{thresh}_\text{EM}$. We want to choose $\mathcal{L}^\text{thresh}_\text{EM}$ so that $\mathcal{L}^n_\text{EM}$ can vary in a suitable range.
If we already know the $\mathcal{L}^\text{max}_\text{EM}$ beforehand, for instance the normalized transmission in a waveguide, we can choose $\mathcal{L}^\text{thresh}_\text{EM}$ so, that $\mathcal{L}^n_\text{EM}=0$ for $\mathcal{L}_\text{EM} = \mathcal{L}^\text{thresh}_\text{EM}$ initially.
If we have no prior knowledge about what constitutes a good value for $\mathcal{L}^\text{thresh}_\text{EM}$, we usually have to do a full optimization without the heat solver to determine a good guess for $\mathcal{L}^\text{thresh}_\text{EM}$. Then we use $\mathcal{L}^\text{max}_\text{EM}$ to tune the initial $\mathcal{L}^n_\text{EM}$. Choosing an initial value of $\mathcal{L}^n_\text{EM}=1$ tends to be a good starting point.

All simulations are done using a single NVIDIA A100 GPU and run for a maximum of 120 optimization steps. The optimization is split into five stages with 20 optimization steps each, or until convergence (absolute change of less than $10^{-4}$ or relative change of less than $10^{-5}$ per iteration), during which we gradually increase $\beta$ from $1$ to $30$ with every stage. Then, a final stage is done with $\beta=30$, and we turn on the binarization penalty.
We discretize our spatial extent using $\SI{40}{px}$ per micrometer.
Our initial design is set to be $\rho=0.5$ everywhere inside the design region.
One full optimization takes roughly half a day to two days, depending on the simulation size and number of sources.

\section{Results} \label{Results}
We showcase the outlined design pipeline with two examples. First, we discuss a focusing element. Second, we discuss a specific waveguide coupler.

\subsection{Focusing element}
Our first example is a focusing element with an operational wavelength of $\SI{1.55}{\upmu m}$. A permittivity of $\varepsilon_r=2.25$ characterizes the written polymer. Maximizing the normalized electromagnetic field at the focal point for a given illumination achieves the desired focusing behavior. Such a figure of merit can be expressed as
\begin{equation}
    \mathcal{L}_{\text{EM}} = \frac{|E_x(x_f, y_f, z_f)|^2}{\int_\mathbb{S} |E_x(x, y, z)|^2 \,\,\,\text{d}x\text{d}y\text{d}z}\,,
\end{equation}
where $E_x(x, y, z)$ is the electric field component in the $x$-direction and $(x_f, y_f, z_f)$ is the focal point. The other electric field components are negligible because the incident field is an $x$-polarized plane wave propagating in the $+z$-direction. The volume within which we optimize the material distribution is a cuboid with side lengths of 4~$\upmu$m~$\times$4~$\upmu$m~$\times$2~$\upmu$m, resulting in roughly $2$ million parameters for our chosen resolution. The full simulation domain has a size of $\SI{5}{\upmu m}\times\SI{5}{\upmu m}\times\SI{6}{\upmu}m$ with the design region placed at the center. Additionally, we use PMLs with width $\SI{0.5}{\upmu m}$ to avoid any unphysical scattering effects. We also pad our design region with material in the $x$-$y$-direction, creating a "frame". This padding acts as a base for fabrication and supports mechanical stability. So we want to connect our heat sinks for the material to said frame. The heat sinks for the void are placed on the top and on the bottom of the design in the $z$-direction. The focal point is chosen to be $\SI{1}{\upmu m}$ behind the terminating interface on the optical axis.

\begin{figure*}
    \centering
    \includegraphics[width=\linewidth]{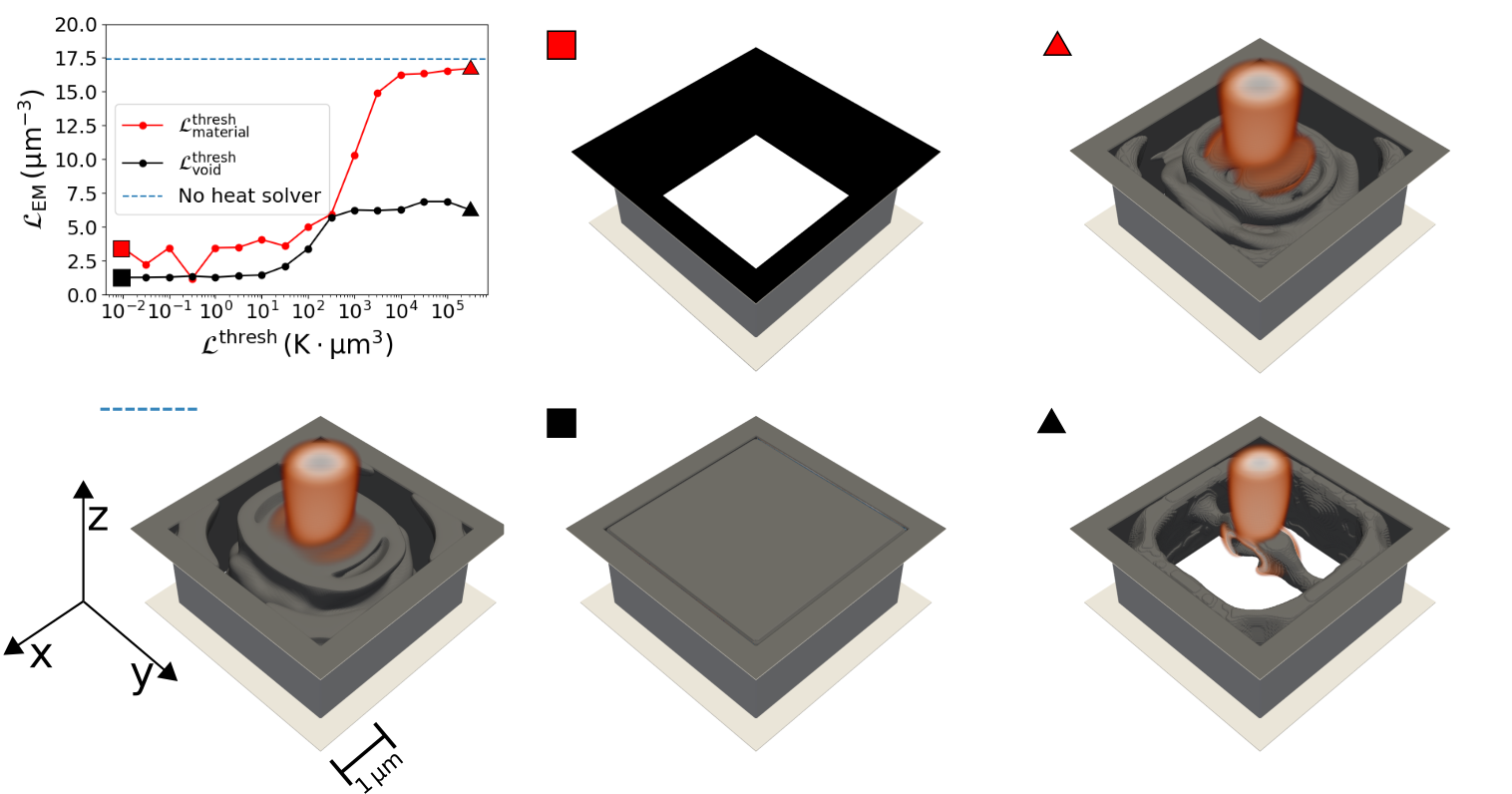}
    \caption{Different focusing devices designed with structural integrity. A parameter sweep is done by varying the threshold value for $\mathcal{L}^\text{thresh}_\text{material}$ while $\mathcal{L}^\text{threshold}_\text{void}$ is constant (red) and by varying the threshold value for $\mathcal{L}^\text{thresh}_\text{void}$ while $\mathcal{L}^\text{thresh}_\text{material}$ is constant (black). Their performance as focusing devices, evaluated using $\mathcal{L}_\text{EM}$ is compared to a device designed by maximizing only $\mathcal{L}_\text{EM}$ (blue dotted line). Some selected optimized devices are also shown, where the markers (colored square or triangle) indicate the corresponding threshold values. The displayed devices show the material structure and the field amplitude above a cut-off value. We also display the device optimized for its optical performance only. It can be seen that this device does not possess structural integrity (see supplementary material for a camera angle that better shows this). The best-performing design with structural integrity (red triangle) can be seen in Fig.~\ref{fig:best_lens} in detail.}
    \label{fig:connectivity_lens}
\end{figure*}
\begin{figure}
    \centering
    \includegraphics[width=\linewidth]{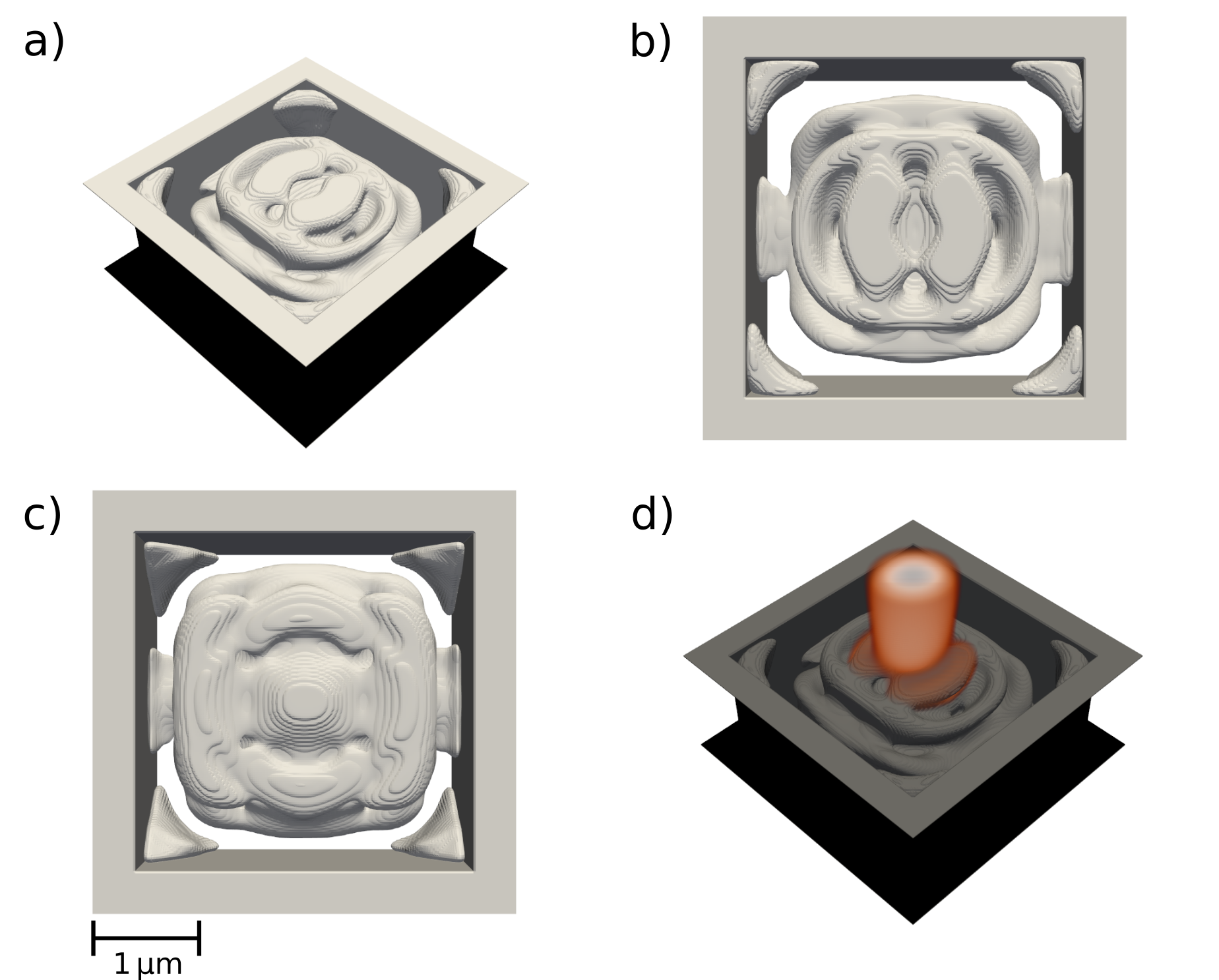}
    \caption{Best performing focusing device with enforced structural integrity, shown from different angles: Tilted (a), top view (b), and side view (c). The electric field distribution $|E_x|$ is shown in (d).}
    \label{fig:best_lens}
\end{figure}
To get insights into the optimization process, we perform two parameter sweeps. In both parameter sweeps we use $\mathcal{L}_\text{EM}$ as a measure of the performance, as we are mostly interested in how a device (with structural integrity) optically performs in comparison to a structure optimized exclusively for its optical functionality.

First, we use a constant value for $\mathcal{L}^\text{thresh}_\text{material}$,  and optimize several devices for different values of $\mathcal{L}^\text{thresh}_\text{void}$.
We select our threshold value such that (given the initial distribution of $\rho=0.5$ everywhere in the design region) the initial value of $\mathcal{L}_\text{material}^n$ is approximately $0.5$.
Second, we do the same by fixing sweeping through $\mathcal{L}^\text{thresh}_\text{material}$ whilst initializing $\mathcal{L}_\text{void}^n \approx 0.5$ in the same way. 
The results of the parameter sweeps can be seen in Fig.~\ref{fig:connectivity_lens}. The figure shows the electromagnetic figure of merit ultimately obtained for each optimized device. As an orientation, we also plot the electromagnetic figure of merit for an optimized device that was not designed for structural integrity. This figure of merit (blue dashed line in Fig.~\ref{fig:connectivity_lens}) serves as an indication of the largest possible value that is achievable when optimizing only the electromagnetic response. 

In addition, Fig.~\ref{fig:connectivity_lens} shows various focusing devices where different values of $\mathcal{L}^\text{thresh}_\text{heat}$ were chosen, as well as the focusing device that was optimized by maximizing only $\mathcal{L}_\text{EM}$. Notably, all designs using an auxiliary heat-dissipation solver possess structural integrity in the material and the void. In contrast, the device that considered only the electromagnetic response does not.

When looking into the results, we observe that if we choose the threshold value too small, the optimization process ends up in structures with $\rho = 0$ or $\rho = 1$ everywhere in the design region. This can be seen clearly in the exemplary devices with the red and black squares, where the optimized structure consists of only a void (red square) or only material (black square). This happens because normalizing to a tiny threshold value results in a thermal objective function that has a much higher value than the electromagnetic objective function by orders of magnitude. Since this happens in the linear regime of the \textit{softplus} function, the gradient will also point towards a solution which optimizes the thermal objective over the electromagnetic objective. For example, when $\mathcal{L}_\text{material}^\text{thresh} \ll 1$, we obtain $\mathcal{L}^n_\text{heat}\gg 1$, while $\mathcal{L}^n_\text{EM}\approx 1$. The optimizer strongly prioritizes minimizing $\mathcal{L}_\text{material}^n$, effectively disregarding $\mathcal{L}_\text{EM}^n$ (and $\mathcal{L}_\text{void}^n$). As a result, the optimization converges to the trivial solution $\rho = 0$ everywhere ({\it i.e.}, no material and therefore no heat sources).
The opposite applies when $\mathcal{L}_\text{void}^\text{thresh} \ll 1$. The optimization fills the entire design region with material ($\rho=1$ everywhere), resulting in no heat generated by the void. Again, this is just a consequence of the fact that if there are no voids, no spatial domain acts as a heat source in the void-connectivity simulation, and thus the temperature stays low everywhere. But of course, these trivial solutions that only satisfy a single thermal sub-objective, whilst disregarding the EM functionality, are not what we want. 
Furthermore, the optimization gets stuck in the local minimum of the trivial solution if it converges towards it initially, even if better local minima are available (see supplementary material).

Increasing the threshold leads to non-trivial structures which can function as proper focusing devices. Interestingly, it seems to be possible to increase the threshold value for either $\mathcal{L}^\text{thresh}_\text{material}$ or $\mathcal{L}^\text{thresh}_\text{void}$ to high values, without losing the structural integrity of material and void. Examples of such optimized devices are also shown in Fig.~\ref{fig:connectivity_lens}, and they are indicated with the red and black triangle, respectively. However, that observation is most likely problem-specific. In the case of choosing high $\mathcal{L}^\text{thresh}_\text{material}$, it is even possible to find designs that are almost as good as a design that was optimized by maximizing only $\mathcal{L}_\text{EM}$. 

In contrast, when considering only $\mathcal{L}_\text{EM}$ in the optimization, the resulting structure is free-floating. We can achieve structural integrity only after adding the auxiliary thermal solvers. The best performing device with structural integrity can be seen in more detail in Fig.~\ref{fig:best_lens}. The electromagnetic object function reaches a value of $\mathcal{L}_\text{EM}=16.7\,\upmu m^{-3}$. Compared to a value of $\mathcal{L}_\text{EM}=17.4\,\upmu m^{-3}$ for the design optimized only for its optical functionality, we can conclude that the enforcement of structural integrity led only to a minor degradation in optical performance. However, we have achieved a fully connected, self-sustaining structure free of voids.

\subsection{Waveguide Coupler}
The second design is a waveguide coupler. We consider two incoming waveguides for two different wavelengths $\lambda_1 = \SI{700}{nm}$ and $\lambda_2 = \SI{600}{nm}$. The waveguides are made from a polymer characterized by a permittivity of $\varepsilon_r=2.25$. Each waveguide couples to an outgoing waveguide rotated by $90^\circ$. We place our heat sinks for the material at the waveguide ports, which is where we are certain that material is required for the optical performance. The heat sinks for the void are placed on the boundaries of the design domain except where the waveguides are. Our design region is defined by a cube with sidelength $\SI{2}{\upmu m}$, resulting in around half a million parameters for our chosen resolution. The full simulation domain has a size of $\SI{3}{\upmu m}\times\SI{3}{\upmu m}\times{6}{\upmu m}$ with the design region at its center. We use PMLs with width $\SI{0.5}{\upmu m}$. As a figure of merit, we choose to maximize the intensity of each field in their respective target waveguides
\begin{align}
\left(\mathcal{L}_\text{EM}\right)^2 &= \frac{\int_{\mathcal{W}_1} |E^1_x(x, y, z_0)|^2\,\,\,\text{d}x\text{d}y}{\int_\mathbb{S}|E^1_x(x, y, z)|^2\,\,\,\text{d}x\text{d}y\text{d}z} \\
&+ \frac{\int_{\mathcal{W}_2}|E^2_x(x, y, z_0)|^2\,\,\,\text{d}x\text{d}y}{\int_\mathbb{S}|E^2_x(x, y, z)|^2\,\,\,\text{d}x\text{d}y\text{d}z},\nonumber
\end{align}
where $E^1_x(x, y, z)$ and $E^2_x(x, y, z)$ are the $x$-component of the electric field for wavelength $\lambda_1$ and $\lambda_2$. In each waveguide, an $x$-polarized $\mathrm{TE}_0$-waveguide mode is propagating, which serves as the illumination. $\mathcal{W}_1$ and $\mathcal{W}_2$ are the cross-section of the outgoing waveguides where each wavelength should couple to. Each waveguide is $\SI{2}{\upmu m}$ long and has a sidelength of $\SI{0.5}{\upmu m}$ and is also displaced by $\SI{0.75}{\upmu m}$ from the center in their respective direction. 
The fields are evaluated inside the outgoing waveguides.
\begin{figure*}
    \centering
    \includegraphics[width=\linewidth]{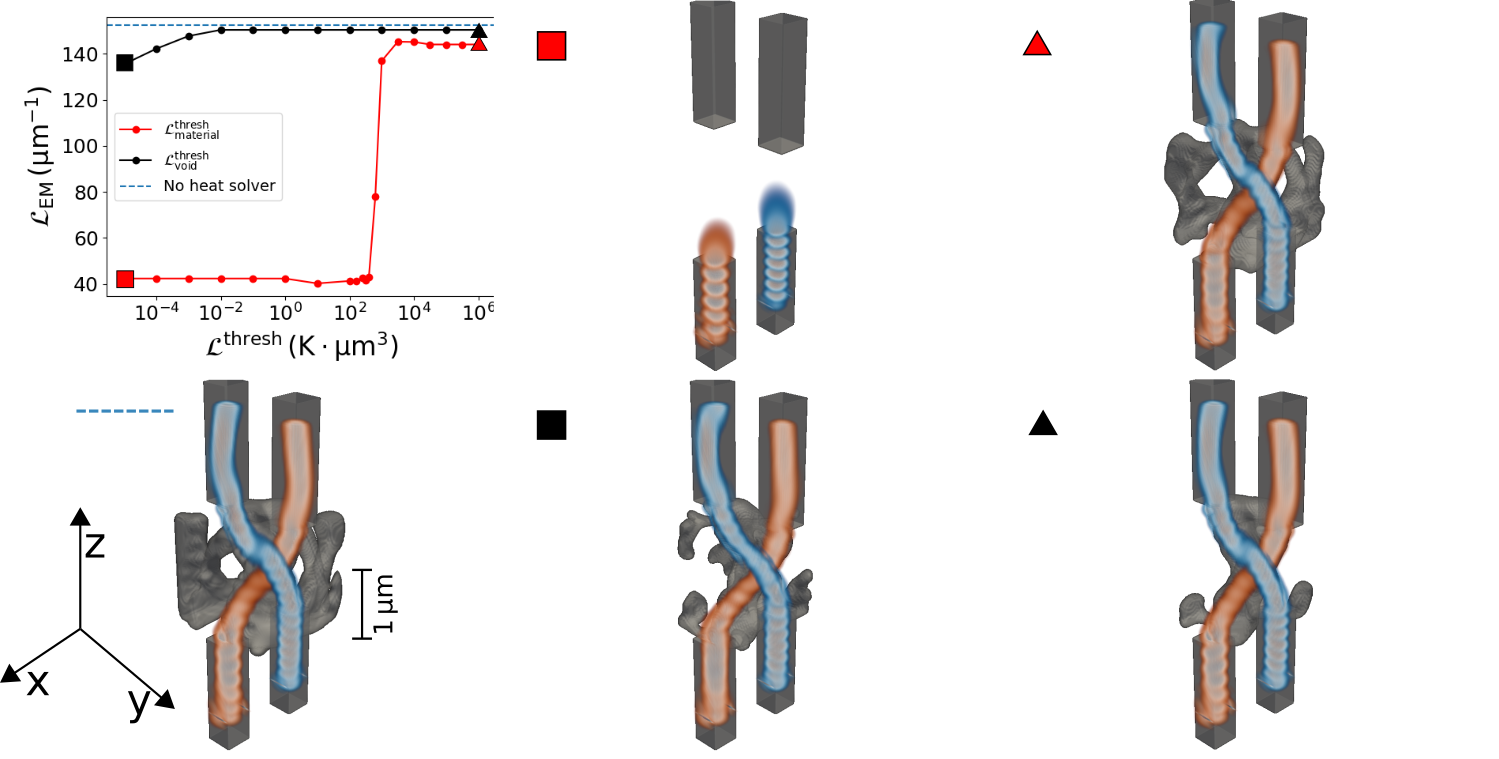}
    \caption{Different waveguide couplers designed with structural integrity. A parameter sweep is done by varying the threshold value for $\mathcal{L}^\text{thresh}_\text{material}$ while $\mathcal{L}^\text{threshold}_\text{void}$ is constant (red) and by varying the threshold value for $\mathcal{L}^\text{thresh}_\text{void}$ while $\mathcal{L}^\text{thresh}_\text{material}$ is constant (black). Their performance as waveguide couplers evaluated using $\mathcal{L}_\text{EM}$ is compared to a waveguide coupler designed by maximizing only $\mathcal{L}_\text{EM}$ (blue dotted line). Some selected optimized devices are also shown, where the markers (colored square or triangle) indicate the corresponding threshold values. The displayed devices
    show the material structure and the field amplitude above a cut-off value. We also display the device optimized for its optical performance
    only. It can be seen that this device does not possess structural integrity (see supplementary material for a camera angle that better shows this). The best-performing design with structural integrity (red triangle) can be seen in Fig.~\ref{fig:best_cross} in detail.}
    \label{fig:connectivity_cross}
\end{figure*}
\begin{figure}
    \centering
    \includegraphics[width=\linewidth]{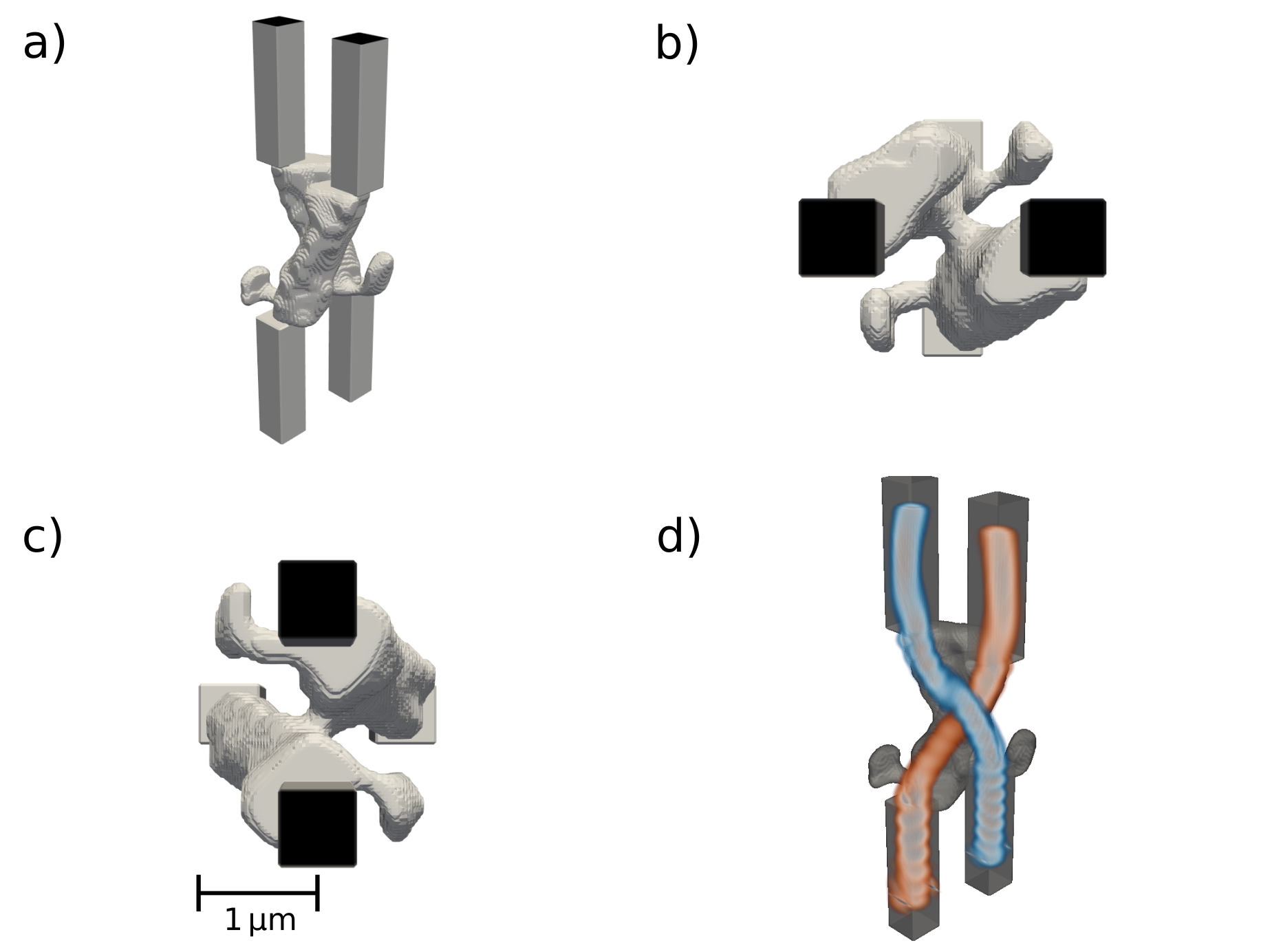}
    \caption{Best performing waveguide crossings with enforced structural integrity, shown from different angles: Tilted (a), top view (b), and side view (c). The electric field distributions $|E^1_x|$ and $|E^2_x|$ with the device are shown in (d).}
    \label{fig:best_cross}
\end{figure}

While the optimization that exclusively optimizes the electromagnetic response already favors structures that mostly possess structural integrity, the final designs may still contain free-floating artifacts and cavities.
As with the focusing device, we sweep through both $\mathcal{L}^\text{thresh}_\text{material}$ and $\mathcal{L}^\text{thresh}_\text{void}$, while fixing the other.
In contrast to the focusing device, we enforce a higher degree of connectivity by setting the threshold values such that the initial values of $\mathcal{L}^n_\text{material}$ and $\mathcal{L}^n_\text{void}$ are approximately 1 (twice as large as for the focusing device).
Lower values still tend to produce free-floating artifacts, which we want to avoid. This is most likely due to the stepwise binarization, which can produce "islands" that are initially connected with non-binarized values, and their connections are slowly eroded.

The results of the parameter sweep can be seen in Fig.~\ref{fig:connectivity_cross}.
As with the focusing device, choosing the value for $\mathcal{L}^\text{thresh}_\text{material}$ too low results in structures where $\rho=0$ almost everywhere in the design region. Therefore, as there is no structure which connects the input with the output waveguide, the electromagnetic objective function is rather low.
There are non-trivial structures that appear for low threshold values of $\mathcal{L}^\text{thresh}_\text{material}$ during the parameter sweep. 
These non-trivial solutions appear due to the placement of the heat sinks, as both input and output waveguides are heat sinks, resulting in solutions which do not connect the waveguides but still minimize the figure of merit with respect to the heat and influence the light in a very non-efficient manner.
Since waveguide-like structures tend to be the optimal solution to the problem, we will end up with fully connected structures for higher threshold values $\mathcal{L}^\text{thresh}_\text{material}$.
At the transition, the bumps begin to elongate towards one another, forming rudimentary guiding paths that outperform the trivial solution but still do not fully connect the input waveguides to the output waveguides. Only once the waveguide ports are connected we can actually guide the light from the input waveguides to the output waveguides in an efficient manner.

The behavior is different when changing $\mathcal{L}^\text{thresh}_\text{void}$. 
Unlike in the case of the focusing element, we do not encounter a solution with $\rho=1$ everywhere in the design region.
The heat sinks for the void make up the majority of the surface of the design region, allowing the heat to be dissipated more easily than for the material. This, in turn, means that the threshold values required for trivial solutions are magnitudes lower than the range we looked at.
Using a high value for $\mathcal{L}^\text{thresh}_\text{void}$ slightly improves the performance while also using less material in the final design. The best-performing design can be seen in Fig.~\ref{fig:best_cross}. This design can reach $\mathcal{L}_\text{EM}=\SI{150}{\upmu m^{-1}}$ in comparison to $\mathcal{L}_\text{EM}=152\,\upmu m^{-1}$ for a design which was optimized by maximizing only $\mathcal{L}_\text{EM}$.
Therefore, by using an auxiliary heat-dissipation simulation, we can remove free-floating islands and cavities while reducing the required material, all with minimal impact on the optical performance.

Interestingly, both with and without the auxiliary heat solver, we find solutions that are not just two waveguides connecting the waveguide ports.
Instead, we get structures with additional features. These additional features help to guide the light through the curved waveguides by catching the field which is propagating outside of the waveguides \cite{10.1117/12.3014421}. This leads to an increased performance, as even the tail end of the electromagnetic waves can be captured.
Once we enforce structural integrity using our heat solver, we favor solutions with less material. The resulting less bulky structures still retain some of the additional features which catch the evanescent field outside the central waveguides, which seemingly still help to improve the performance of our waveguides.
This phenomenon is similar to additional features appearing in 2D waveguide bends \cite{schubertInverseDesignPhotonic2022, irfan2024ultra}, where resonator-like structures appear to improve the performance of the waveguide bend.

\section{Discussion}\label{Discussion}

This work introduces a robust methodology to ensure the structural integrity of 3D-printed nanophotonic devices through the integration of an auxiliary heat-diffusion solver within a gradient-based topology optimization framework. Our approach not only prevents disconnected or structurally unstable designs but also preserves the desired optical functionalities with minimal performance trade-offs.

By leveraging the heat-diffusion solver as a fictitious physics-based soft-constraint, we effectively enforce connectivity in both material and void domains. The use of renormalised figures of merit and the \textit{softplus} function ensures that thermal optimisation ceases once structural integrity is achieved, allowing the optimizer to focus on the optical performance. In combination with the iterative binarization strategy, our approach results in designs that are directly fabricable, offering a streamlined pipeline for 3D nanoprinting applications.

The proposed framework was demonstrated through the design of two devices—a focusing element and a waveguide crossing. These case studies validate the effectiveness of our method in achieving structurally robust and optically performant devices. In both cases, the devices with structural integrity performed only marginally worse than those optimized for their optical performance only.

This study lays the foundation for a broader application of physics-based constraints in the design of complex 3D nanostructures. Future work could explore extensions of this methodology to include additional auxiliary (or indeed real) physical constraints. Further investigation into adaptive weighting schemes and thresholding strategies could also enhance the robustness and versatility of the optimization process. By addressing the longstanding challenge of structural integrity in 3D nanophotonic design, this work represents a significant step forward in enabling practical, high-performance devices for real-world applications.

\begin{acknowledgement}
This work was performed on the HoreKa supercomputer funded by the Ministry of Science, Re-
search and the Arts Baden-Württemberg and by the Federal Ministry of Education and Research.
\end{acknowledgement}
\begin{funding}
O.K. and C.R. acknowledge support from the German Research Foundation within the Excellence Cluster 3D Matter Made to Order (EXC 2082/1 under project number 390761711) and by the Carl Zeiss Foundation.
T.J.S. acknowledges funding from the Alexander von Humboldt Foundation.
\end{funding}
\begin{authorcontributions}
All authors have accepted responsibility for the entire content of this manuscript and consented to its submission to the journal, reviewed all the results and approved the final version of the manuscript. C.R., T.J.S. and Y.A. conceived the idea and surpervised the research. R.N.H. has done initial investigations for 2D, O.K. wrote the code and designed the devices in 3D. O.K. wrote the manuscript. C.R. and T.J.S. revised the paper.
\end{authorcontributions}

\begin{conflictofinterest}
Authors state no conflict of interest.
\end{conflictofinterest}
\begin{dataavailabilitystatement}
The code to reproduce the datasets analyzed during the current study are available in the GitHub repository, \url{https://github.com/OlloKuster/Structural_Integrity_3D/tree/main}.
\end{dataavailabilitystatement}
\vspace*{-22pt}

\bibliographystyle{unsrt}
\nocite{*}
\bibliography{article}

\clearpage

\title{Supplementary Material}
\maketitle

\section{Optimized structures without a heat solver}
Here we show the respective optimization results without using an auxiliary heat solver.
In both cases our figure of merit is $\mathcal{L} = -\mathcal{L}_\text{EM}$, only optimizing for the optical performance of the designs. Other than that, the setups are exactly the same as described in the main text

\subsection{Focusing Device}
\begin{figure}[h]
    \centering
    \includegraphics[width=\linewidth]{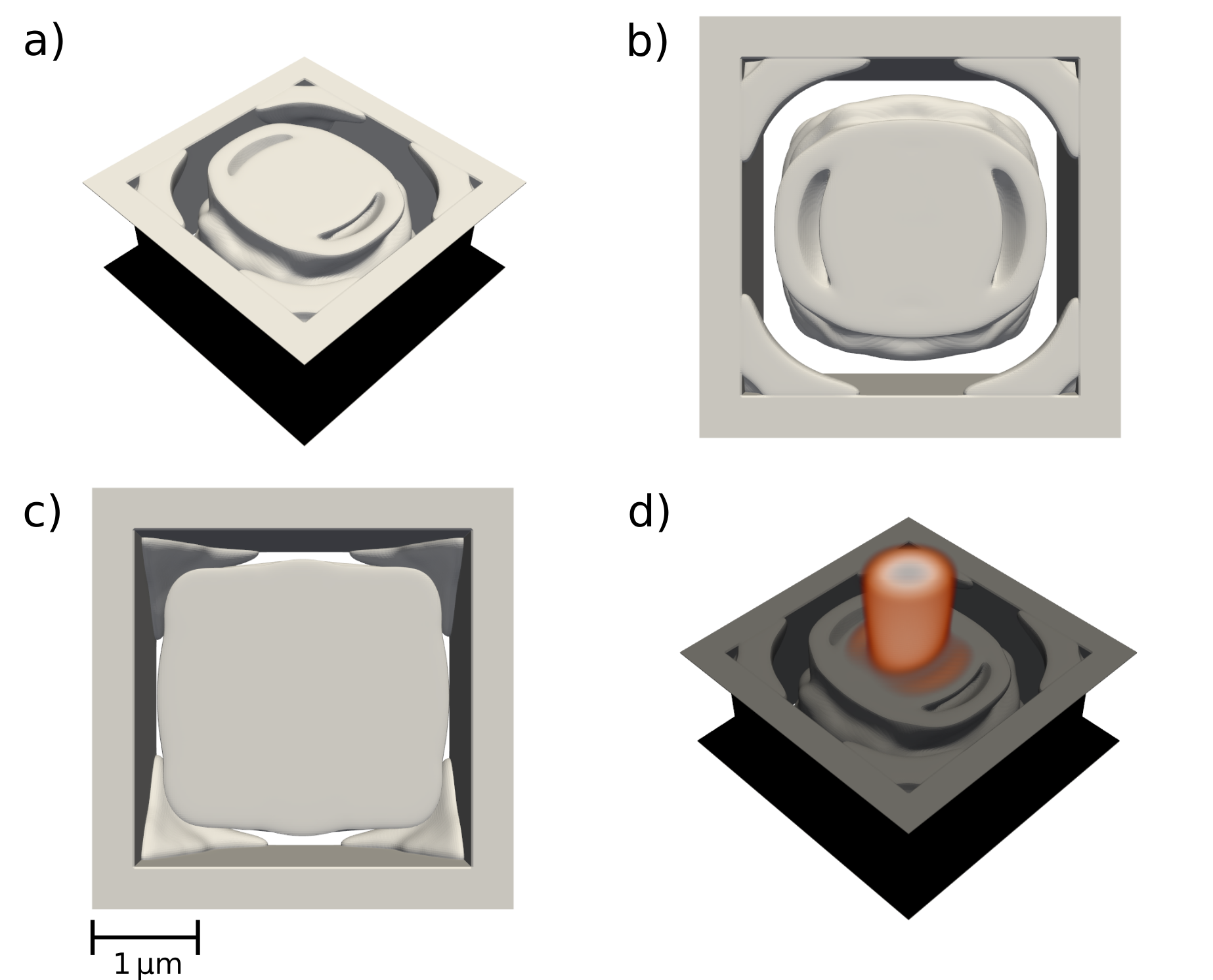}
    \caption{a)-c) optimized focusing device without the heat constraints from different angles. d) shows the optimized design with the enhanced electric field distribution.}
    \label{fig:base_lens}
\end{figure}

Figure \ref{fig:base_lens} shows the optimized design. It can be clearly seen, that the center of the design is free floating, without any additional support structures. What cannot be seen is, that the design also contains cavities. These cavities can be seen in the animations provided in the additional supplementary material.

\subsection{Waveguide Coupler}
\begin{figure}[h]
    \centering
    \includegraphics[width=\linewidth]{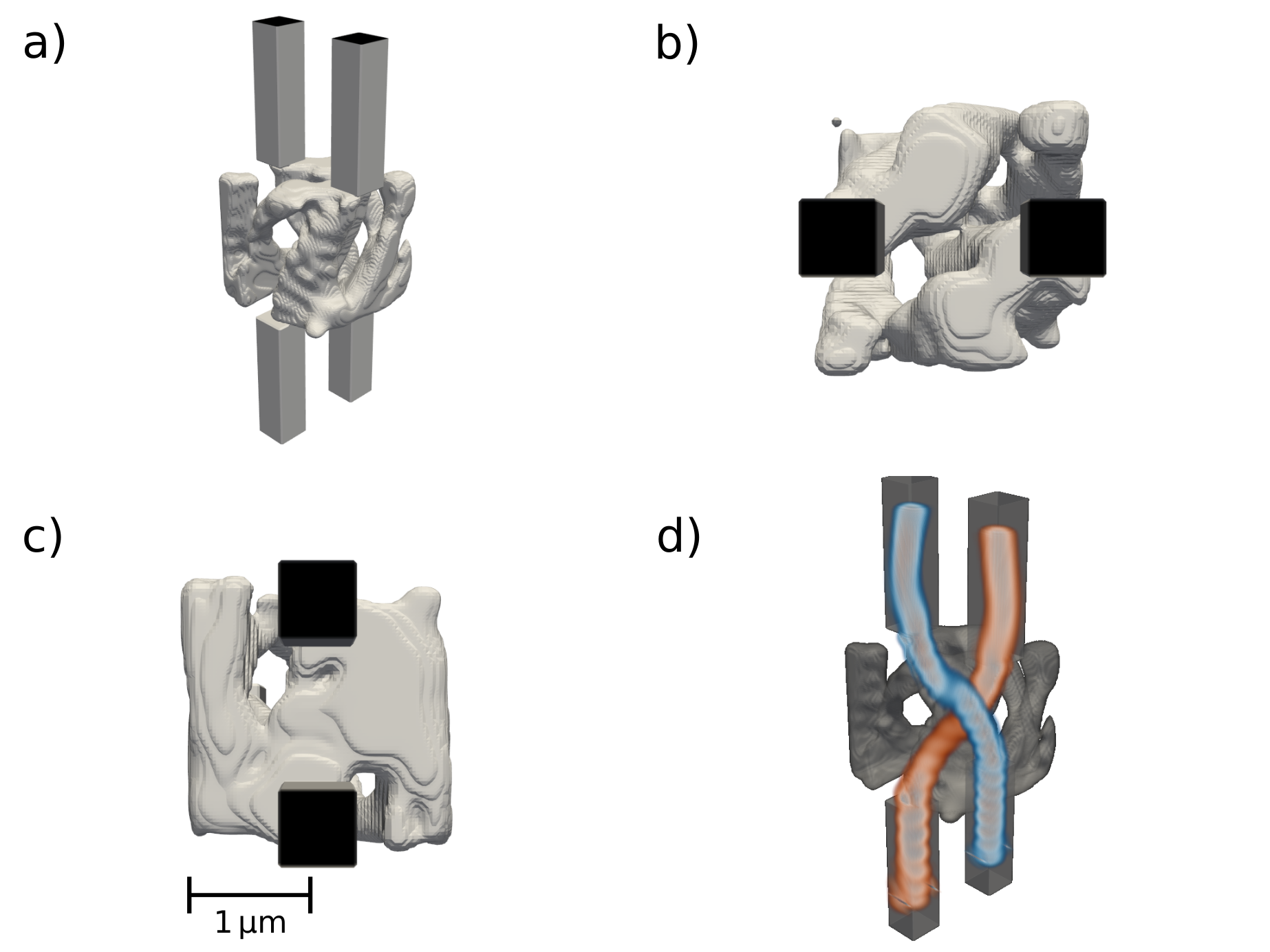}
    \caption{a)-c) optimized waveguide coupling device without the heat constraints from different angles. d) shows the optimized design with the enhanced electric field distribution.}
    \label{fig:base_cross}
\end{figure}

Figure \ref{fig:base_cross} shows the optimized design. While the majority of the design is structurally integral, some free floating parts appear. The design does not contain any cavities, which can also be seen in the animations provided in the supplementary material.
It should still be noted, that this design uses more material than a design which uses the auxiliary heat solver, while achieving minimally improved performance.

\section{Animations}
We provide several animations, which show the optimized designs with and without the auxiliary heat solver. These can be found under \url{https://github.com/OlloKuster/Structural_Integrity_3D/tree/main/Animations}. 16 different animations are shown. First 2 for two directions are given and denoted by "\_x/y". Then we have simulations for the "material" and "void" each and for "opt" and not, meaning optimized using an auxiliary heat solver and without an auxiliary heat solver. Lastly, we provide "heat" plots, which show $u(x, y, z)$ in addition to the structure.
Note, that the material/void plots are done using contour plots. The designs themselves are not hollow, but for better visualization, we use the contour plots.

\section{Probing the local minima}
To get a deeper understanding of how the figure of merit works, we look into the behavior of the optimization. For simplicity, we only look at the focusing device, as the transition between the trivial structures and non-trivial structures is seen more easily in this example.

\begin{figure}
    \centering
    \includegraphics[width=\linewidth]{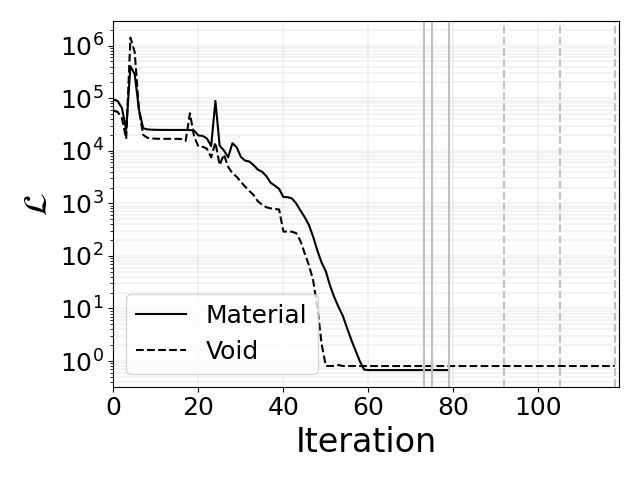}
    \caption{Two optimizations of a focusing device. The optimization was conducted until convergence for a low threshold value, leading to a trivial solution (roughly after 60 iterations). Then the optimization is continued from the previously found local minimum by increasing the threshold values. Each vertical line indicates an increase of the threshold value by two orders of magnitudes. We see that the optimizer stays in the initially found solution and is not able to escape its local minima and stays at the trivial solutions.}
    \label{fig:evolution_lens}
\end{figure}

First, we use a very small threshold value $\mathcal{L}^\text{thresh}_\text{heat}$. The optimizer finds a trivial structure that consists of only material or only void (see red and black square in Fig.~2 of the main manuscript respectively). Once we have converged (which happens roughly after 60 iterations), we increase the threshold value $\mathcal{L}^\text{thresh}_\text{heat}$ and optimize the device from that initially obtained solution. We repeat this process a few times. The results of these optimizations can be seen in Fig.~\ref{fig:evolution_lens}. The vertical lines mark the points where we increased the threshold value by two orders of magnitudes.
After each increase, we kept the optimization going until convergence (as stated in the main text).
As we do know that there are non-trivial solutions for higher threshold values, i.e., they were obtained by optimizing the designs for each given threshold value from scratch, we know that their respective local minima do exist for the figure of merit.
Nonetheless, the optimization is unable to reach local minima which represent a functional optical device and is stuck in the local minimum it converged towards initially.
While the optimization shown in \autoref{fig:evolution_lens} is able to reach values of $\mathcal{L}\approx0.8$, the optimization mostly optimized for the thermal performance where we start with $\mathcal{L}^n_\text{heat}\approx 6000$, whereas $\mathcal{L}^n_\text{EM}\approx 0.6$ for for the initial simulation. Ideally, we want to reach $\mathcal{L}\le 0.2$ for a well functioning, structurally connected device. By optimizing for the thermal performance first, the electromagnetic sub-objective stays relatively constant throughout the entire optimization, or can get even worse, leading to the trivial solutions shown in the main manuscript.

This suggests that we cannot escape the local minimum once we have converged towards it, even if there is a more suitable one available somewhere else. However, the problem is circumvented by optimizing the devices from scratch for every considered threshold value, as done in the main manuscript.

\begin{figure}
    \centering
    \includegraphics[width=\linewidth]{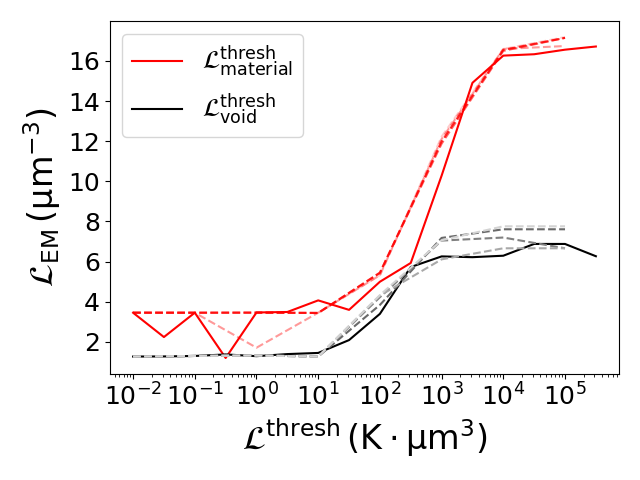}
    \caption{Dependence of $\mathcal{L}_\text{EM}$ with respect to the threshold value of the heat problem. The red curves show the dependence of $\mathcal{L}_\text{EM}$ on $\mathcal{L}_\text{material}^\text{thresh}$ with a fixed void threshold value. The black-gray curves show the dependence of $\mathcal{L}_\text{EM}$ on $\mathcal{L}_\text{void}^\text{thresh}$ with a fixed material threshold value. Various randomly initialized density distributions are shown as dashed lines. The solid lines present the dependence as shown in the main manuscript with a density distribution which was initialized with a uniform distribution.}
    \label{fig:random_init}
\end{figure}

Secondly, we look at which local minima are found with randomly initialized density distributions. 
To do so, we repeat the procedure detailed in the main text, but instead of using a uniform density distribution, we use seeded, randomly initialized density distributions. Figure~\ref{fig:random_init} shows the best found $\mathcal{L}_\text{EM}$ for different threshold values $\mathcal{L}^\text{thresh}_\text{heat}$. Overall, almost no difference between the different randomly initialized and the uniformly initialized distribution can be found.


\section{Resolution of the simulation}
\begin{figure}
    \centering
    \includegraphics[width=\linewidth]{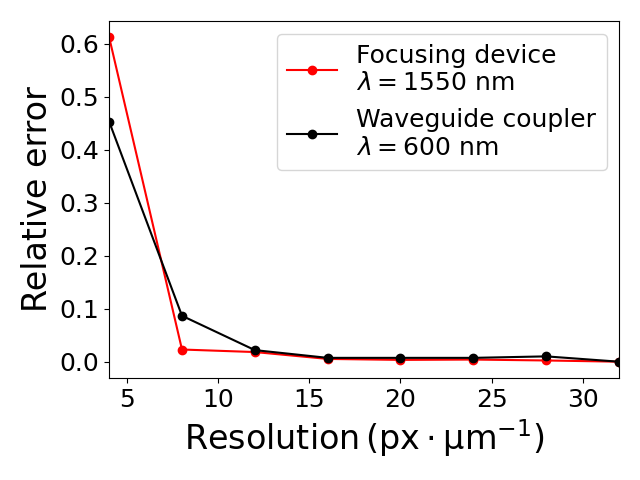}
    \caption{Evaluation of the impact of the spatial resolution in the FDFD simulation on the results. The figure shows the relative error of the electromagnetic objective function of the focusing device and the waveguide coupler as shown in main manuscript depending on the spatial resolution considered in the FDFD simulations. Indeed, we observe convergence already for a resolution of roughly ten pixels per micrometer. This is in agreement with the common expectation of dielectric materials in FDFD for the considered wavelengths.}
    \label{fig:resolution}
\end{figure}

To verify if our chosen spatial resolution in the simulations is sufficient, the simulations are run at a lower resolution with the optimized structures, and we analyze the convergence behavior of the electromagnetic objective function depending on the resolution. The results of these additional simulations can be seen in Fig.~\ref{fig:resolution}, where we changed the spatial resolution in terms of pixel per micrometer in our finite-difference finite-domain (FDFD) simulations. Here, we study the impact of the spatial resolution considering the most optimal focusing device, i.e., the focusing device shown in Fig.~3 of the main manuscript and the optimal waveguide coupler shown in Fig.~5 of the main manuscript.
We evaluated the electromagnetic figure of merit as described in the main text but changed the spatial resolution at which the optical simulations are performed as well as the resolution of the considered device.
We then compare the relative error of the resulting electromagnetic figure of merit
\begin{equation}
    E = \frac{|\mathcal{L}_\text{EM}^\text{opt} - \mathcal{L}_\text{EM}^\text{current}|}{\mathcal{L}_\text{EM}^\text{opt}},
\end{equation}
where $\mathcal{L}^\text{opt}_\text{EM}$ refers to the figure of merit the structure at $\SI{40}{px \cdot \upmu m^{-1}}$ evaluates at and $\mathcal{L}_\text{EM}^\text{current}$ refers to the figure of merit of a structure with the respective lower resolution.
We note, that we only evaluated the waveguide coupler at \SI{600}{nm} wavelength, as a sufficient resolution at \SI{600}{nm} would also be sufficient for \SI{700}{nm}. Additionally, The figure of merit considered in the main manuscript is not independent of the resolution. So to properly analyze the convergence behavior depending on the resolution, we renormalize the electromagnetic figure of merit for the waveguide coupler by the crossectional area of the waveguide $\mathcal{L}^\text{renorm}_\text{EM} = \frac{\mathcal{L}_\text{EM}}{w^2}$, where $w$ is the width of the waveguide. 

Clearly, the resulting error converges already right around $\SI{10}{px\cdot\upmu m^{-1}}$ in both cases. This is not surprising, considering the fact that the polymer materials that are considered in the structure have a rather low permittivity and are on their own optically not that complicated to evaluate.

\section{Non-trivial solutions for the waveguide coupler}
\begin{figure}
    \centering
    \includegraphics[width=\linewidth]{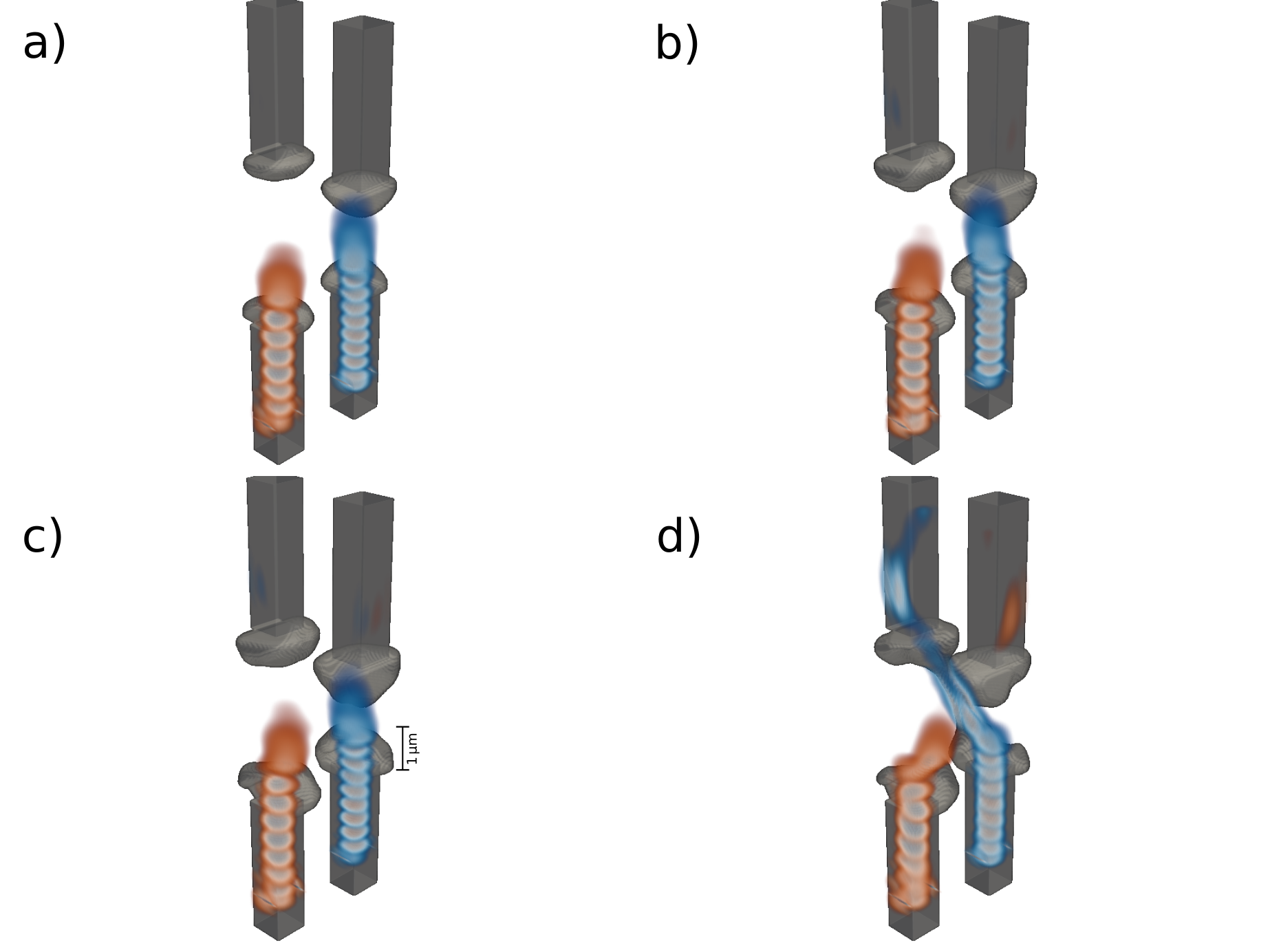}
    \caption{a) - c) Different solutions found for the waveguide coupler which are not completely trivial (e.g. only void) but have little to no influence on the electromagnetic performance. d) shows a device where one of the ingoing waveguides is connected to its outgoing waveguide, but not both. The threshold values used are a) $\mathcal{L}_\text{material}^\text{thresh} = \SI{100}{K\cdot \upmu m^3}$, b) $\mathcal{L}_\text{material}^\text{thresh} = \SI{251}{K\cdot \upmu m^3}$, c) $\mathcal{L}_\text{material}^\text{thresh} = \SI{398}{K\cdot \upmu m^3}$ and d) $\mathcal{L}_\text{material}^\text{thresh} = \SI{631}{K\cdot \upmu m^3}$ and $\mathcal{L}^n_\text{void}=\SI{1}{K\cdot \upmu m^3}$ for all of these examples.}
    \label{fig:bumps}
\end{figure}

Fig.~\ref{fig:bumps} shows solutions to the waveguide coupling problem which do not consist of only void and are thus non-trivial. These solutions appear before and right around the sharp increase in performance seen in Fig.~4 of the main manuscript for the sweep of $\mathcal{L}_\text{material}^\text{thresh}$. Still, these solutions (with the exception of d)) are not able to meaningfully guide the light to the outgoing waveguides. Since the heat sinks sit on all of the four waveguides, structures which do not connect the ingoing to the outgoing waveguide can be found which is why they do not qualify as fully connected in a strict sense. 
\section{Softplus function}
\begin{figure}
    \includegraphics[width=\linewidth]{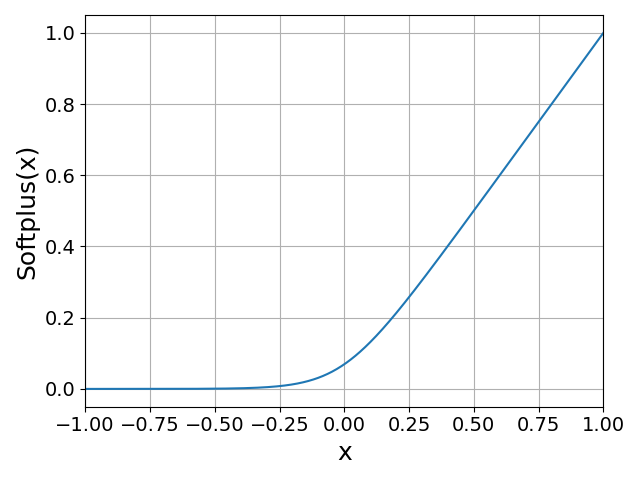}
    \caption{Softplus function}
    \label{fig:softplus}
\end{figure}

Figure \ref{fig:softplus} shows the \textit{softplus} function in the range $[-1, 1]$. Our renormalization is done in a way, that $-1$ denotes the lowest possible value, while the positive range is unbound. We start to interpolate linearly after $x=0.4$ for simplicity.
Values below $x=0$ start to rapidly decay, leading to an optimization procedure which automatically puts emphasis on values which haven't crossed their chosen threshold value.

\clearpage

\end{document}